%% file: report_nips_springer.tex
\documentclass{llncs}
\pdfoutput=1
\usepackage{llncsdoc}
\usepackage[pdftex]{graphicx}
\usepackage[cmex10]{amsmath}
\usepackage{fixltx2e}
\usepackage{url}
\usepackage{stfloats}
\usepackage[utf8]{inputenc}
\usepackage{amssymb}
\usepackage{mathtools}
\usepackage{hyperref}
\usepackage{pgfplots} 
\usepackage{tikz}
\usepackage[font=footnotesize]{subfig} 
\usepackage[autostyle]{csquotes}
\usepackage[
    backend=bibtex,
    style=numeric,
    sorting=none,
    natbib=true,
    url=true, 
    doi=true,
    eprint=false
]{biblatex}
\newlength\Colsep
\usepackage{geometry}
\begin{document}
\author{Tammo Rukat\inst{1} \and Adam Baker\inst{2} \and Andrew Quinn\inst{2} \and Mark Woolrich\inst{2}}
\authorrunning{T. Rukat et al.}
\institute{Department of Statistics, University of Oxford, Oxford, United Kingdom
  \email{tammo.rukat@stats.ox.ac.uk} \and 
  Oxford Centre for Human Brain Activity, Oxford, United Kingdom}
\title{Resting state brain networks from EEG:\\ Hidden Markov states vs. classical microstates}

\maketitle

\begin{abstract}
Functional brain networks exhibit dynamics on the sub-second temporal scale and are often assumed to embody the physiological substrate of cognitive processes.
Here we analyse the temporal and spatial dynamics of these states, as measured by EEG, with a hidden Markov model and compare this approach to classical EEG microstate analysis. We find dominating state lifetimes of 100--150\,ms for both approaches. The state topographies show obvious similarities. However, they also feature distinct spatial and especially temporal properties.
These differences may carry physiological meaningful information originating from patterns in the data that the HMM is able to integrate while the microstate analysis is not. This hypothesis is supported by a consistently high pairwise correlation of the temporal evolution of EEG microstates which is not observed for the HMM states and which seems unlikely to be a good description of the underlying physiology.
However, further investigation is required to determine the robustness and the functional and clinical relevance of EEG HMM states in comparison to EEG microstates. 
\end{abstract}



%
	\newlength\figureheight 
	\newlength\figurewidth

\section{Introduction} \label{sec:introduction}
Temporal correlations in the spontaneous oscillatory activity of spatially distinct neuronal assemblies are a well established phenomenon described as \textit{resting state brain networks} (RSNs). RSNs exhibit functional \cite{Guerra-Carrillo2014} and clinical \cite{Filippini2009,Centeno2014,Lang2014,Ovadia-Caro2014} significance. They have first been identified based on blood-oxygen levels measured through functional MRI \cite{Beckmann2005,Damoiseaux2006}. While fMRI is limited in its temporal resolution and captures only slow oscillations with frequencies below $0.1$\,Hz, it features a high spatial resolution down to $1$\,mm. In contrast, \textit{electroencephalography} (EEG) and \textit{magnetoencephalography} (MEG) are techniques that provide a more direct measure of the electrical activity in the brain \cite{Michel2009,Proudfoot2014}. EEG measures the difference in electric potentials on the scalp and captures high frequency oscillations on the millisecond timescale that is most relevant for the characterisation of cognitive processes. It therefore is a suitable tool to characterise the electrophysiological basis of RSNs~\cite{He2008,Liu2010}.
Notably, the same resting state patterns can be observed across the different time scales of fMRI and M/EEG, which is made conceivable by the dynamics of brain states being scale free across the relevant regime \cite{Linkenkaer-Hansen2001,Kitzbichler2009,VandeVille2010}.

Here, we apply a hidden Markov Model (HMM) \cite{Rezek2002,Vidaurre2015_unp} to the power envelope of the EEG signal in sensor space, in order to identify quasi-stable networks of correlated activation that the signal is likely to have emerged from. This model has previously been applied to resting state MEG power envelopes and identified brain states that appear highly similar to known RSNs~\cite{Baker2014}, featuring rapid fluctuations with state lifetimes of 100--200\,ms.
The aim of this study is to investigate the HMM's potential as an alternative or complementary method to classical EEG microstate analysis and to compare their spatial and temporal characteristics.
In classical microstate analysis, the EEG signal is thought of as a sequence of a limited number of quasi-stable EEG topographies, each defining a microstate~\cite{Koenig1999,Khanna2015}. These are inferred based on the EEG topographies at local maxima of the global field power (GFP), which is given by the sum squared difference between all electrode potentials $V_i$ and the mean potential $\overline{V}$:
$ \mbox{GFP}(t) = (\frac{1}{n} \sum_i V_i(t)-\overline{V}(t))^{\frac{1}{2}} $.

Topographies at the GFP maxima are of particular interest because they feature the highest signal-to-noise ratio \cite{Koenig2005}.
The selected topographies are subject to a clustering procedure that aims to determine a fixed number of states, reflecting typical topographies. Traditionally this is achieved by an iterative procedure \cite{Pascual-Marqui1995}, while K-means clustering or more sophisticated hierarchical clustering methods \cite{Michel2009,Tibshirani2005} have become the current standard.
Classical microstate analysis often limits itself to 4 clusters that have repeatedly been observed to explain most variance in the data. 
Time courses can then be derived under the assumption that the switching between mutually exclusive microstates happens only at GFP peaks. This procedure yields mean state durations of around $100$\,ms \cite{Brodbeck2012}. Microstates have a variety of clinical applications, e.g. in Schizophrenia \cite{Lehmann2005} and Alzheimer's \cite{Nishida2013}.
where duration and switching patterns between the four microstates are connected to the disease state. However, the extent to which they reflect topographies of physiological activation remains unclear. 

\section{Data acquisition, preprocessing and classical microstate analysis} 
\begin{figure}[thp]
\centering
\addtocounter{subfigure}{-1}
\renewcommand{\thesubfigure}{\normalsize{\arabic{subfigure}}}
\subfloat[\label{subfig:ms_topos}\normalsize{Classical EEG microstates}]{%
\renewcommand{\thesubfigure}{\normalsize{\alph{subfigure}}}
\subfloat[]{%
\includegraphics[width=.11\linewidth]{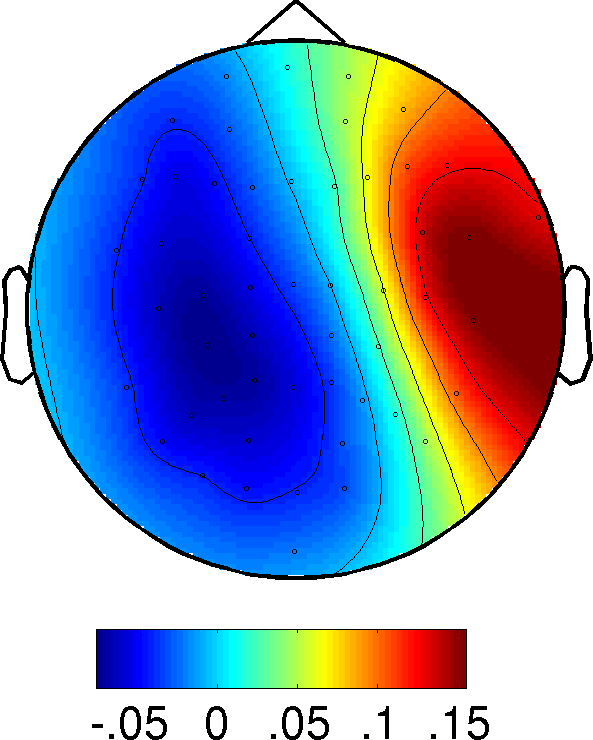}}\vspace{0.01\linewidth}
  \subfloat[]{%
\includegraphics[width=.11\linewidth]{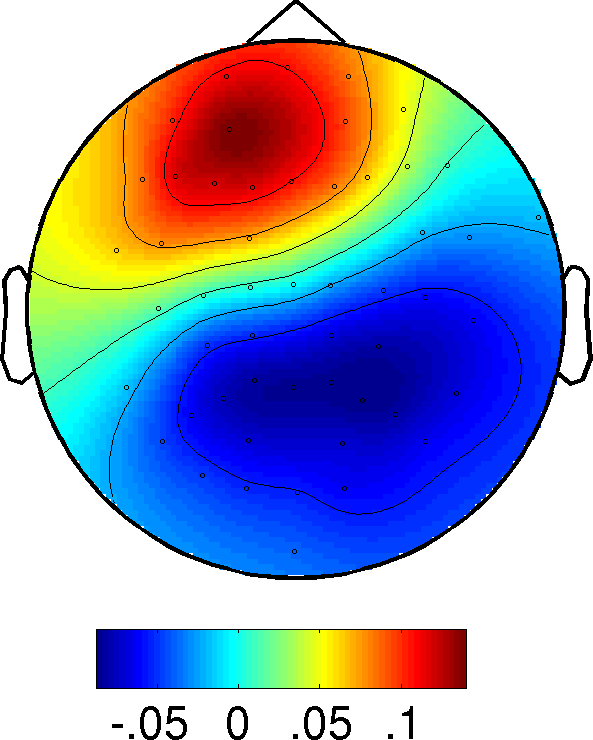}}\vspace{0.01\linewidth} 
  \subfloat[]{%
\includegraphics[width=.11\linewidth]{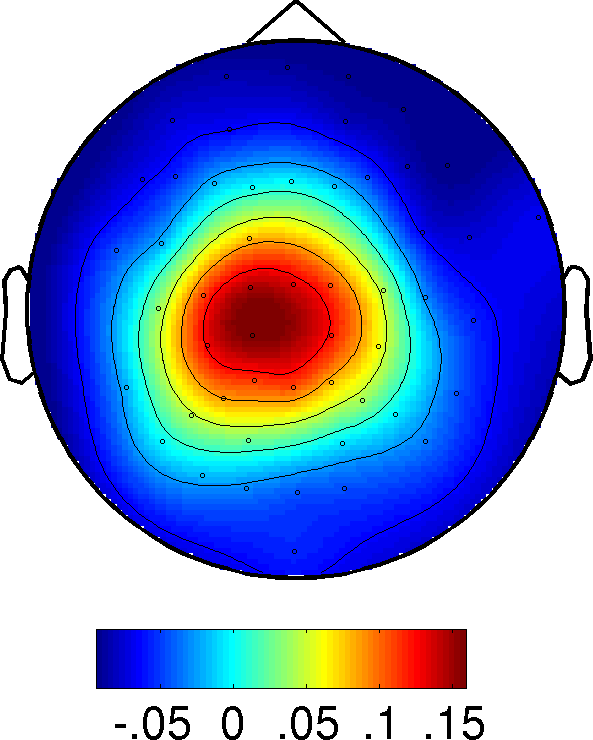}}\vspace{0.01\linewidth}
  \subfloat[]{%
\includegraphics[width=.11\linewidth]{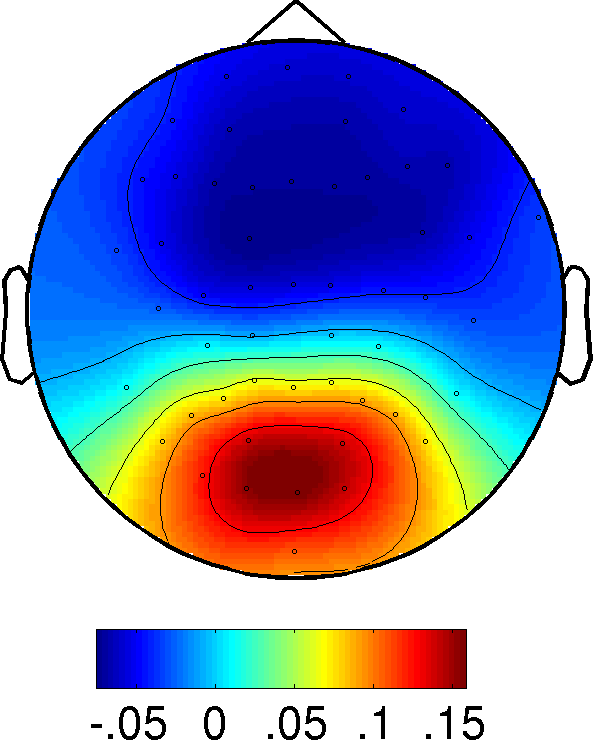}}
\addtocounter{subfigure}{-3}}\;\addtocounter{subfigure}{-2}%
\subfloat[\normalsize{Hidden Markov states}\label{subfig:hmm_topos}]{%
\renewcommand{\thesubfigure}{\normalsize{\alph{subfigure}}}
 \subfloat[]{
\includegraphics[width=.11\linewidth]{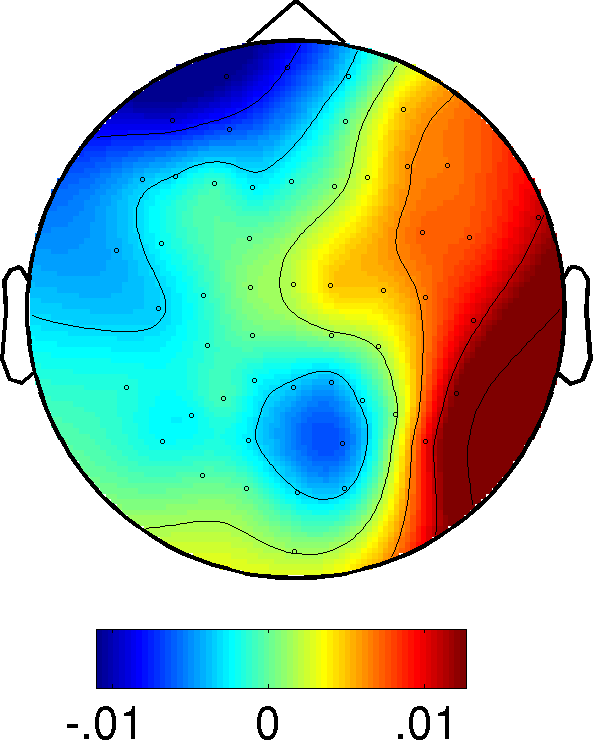}}\vspace{0.01\linewidth}
  \subfloat[]{%
\includegraphics[width=.11\linewidth]{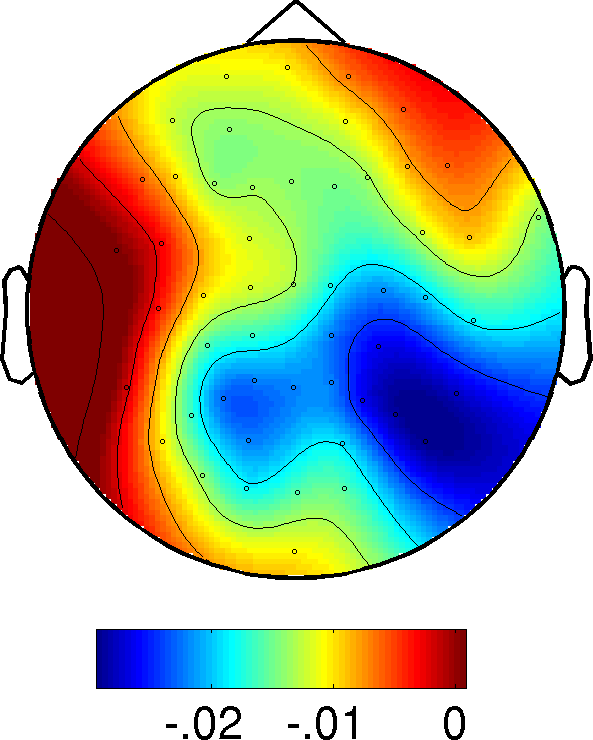}}\vspace{0.01\linewidth}
  \subfloat[]{%
\includegraphics[width=.11\linewidth]{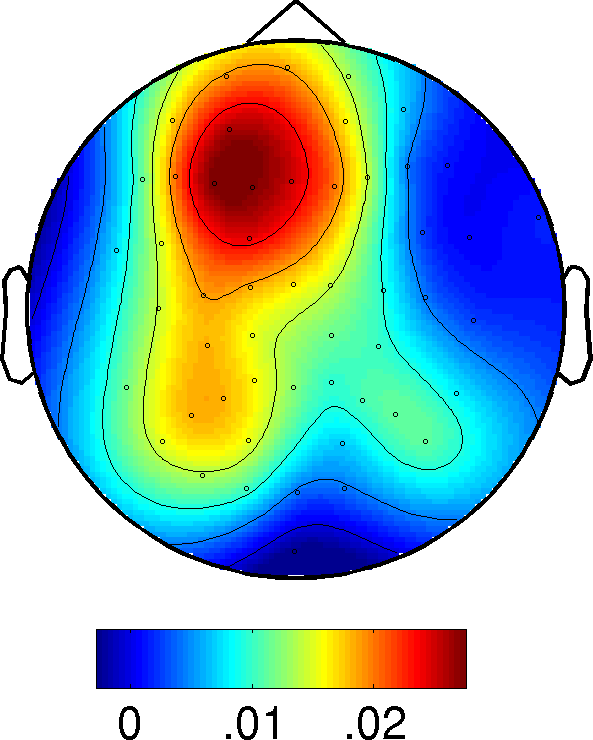}}\vspace{0.01\linewidth}
  \subfloat[]{%
\includegraphics[width=.11\linewidth]{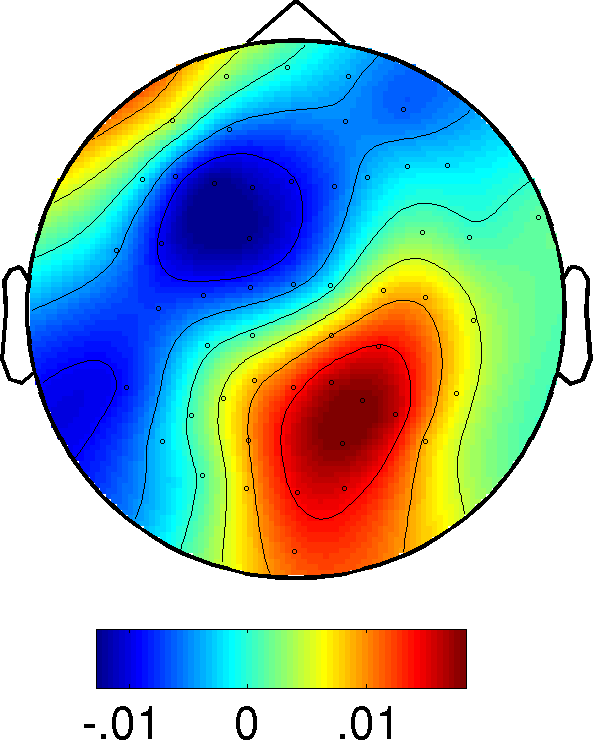}}
\addtocounter{subfigure}{-2}
\renewcommand{\thesubfigure}{\normalsize{\arabic{subfigure}}}
}
\caption{\label{fig:eeg_topos} \textbf{Quasi-stable spatial EEG topographies} -- Based on resting state EEG measurement for 2x10 minutes in 6 subjects. The red-blue colour coding shows opposite potentials. As opposed to the microstates, the range in potentials differs among HMM states. They are separately normalised to facilitate the comparison between modalities.}
\end{figure}

For this study we recorded two times ten minutes resting state EEG data in 6 healthy subjects. Upon identification based on the cardiac signal, the eye blink signal and the signal's kurtosis and frequency spectrum, the data is manually cleaned from apparent artefacts. Subsequently it is decomposed into 150 independent components and band-pass filtered into the 1--40\,Hz band.
EEG Microstates are inferred, based on global field power (GFP) time course, smoothed with a Gaussian kernel with a width of 10 time steps and a standard deviation of 5 time steps. GFP peaks are considered local maxima if all 10 surrounding values are smaller. Upon identification, the peak topographies are subject to k-means clustering with a fixed number of clusters, where the objective function is the within sample correlation for each cluster. The mean distance for each topography from its assigned centroid under variation of the number of states 
does not immediately suggesting a certain number of clusters that is particularly well supported by the data (not shown).
This procedure yields microstates that are shown in Fig.~\ref{fig:eeg_topos}~\subref{subfig:ms_topos} and that appear similar to those found in the literature~\cite{Koenig2002}.

\section{A hidden Markov model for EEG and MEG topographies}
\label{sec:hidden-markov-model}
In contrast to classical microstate analysis, the hidden Markov model, as proposed here, is a generative model that describes the observations that emerge from the rapid switching between quasi-stable topographies with a Gaussian observation model. It promises to be able to capture temporal and spatial dynamics that are more closely related to the underlying brain activity than classical microstate analysis and has been successfully applied to the analysis of MEG RSNs~\cite{Baker2014}. We now briefly outline the model of which a detailed account is given elsewhere~\cite{Vidaurre2015}. 

\subsection{Model derivation}
\label{sec:derivation}
At any given time $t$ the system is in a state $k$ out of fixed number, $K$, of states, denoted $s_t$. Each of these states is associated with a Gaussian observation model that describes the mean and covariance for every data point. With $y_t$ denoting the vector of observation at time $t$ we therefore write:
\begin{align}
  P(y_t|s_t=k,\mu_k,\Sigma_k) \sim \mathcal{N}(\mu_k,\Sigma_k). 
\end{align}
The transition probability between states is Markovian, such that
\begin{align}
 P(s_t=k|s_{t-1}=k',s_{t-2}=k'',\ldots) = P(s_t=k|s_{t-1}=k') = \pi_{k,k'}
\end{align}
where the transition probabilities from state $k$ to $k'$ are described by $K$x$K$ matrix $\pi$. The initial probability to be in state $k$ is given by $\pi_0$. 
The full posterior likelihood is given by:
\begin{align}
 P(y,s,\pi_0,\pi,\mu_k,\Sigma_k) = \prod\limits_t  P(y_t|s_t,\mu_k,\Sigma_k)P(s_t|s_{t-1},\pi)P(\pi_t)P(\pi_0) P(\mu,\Sigma)
\end{align}
Choosing conjugate distributions for the priors, $P(\pi_t)$, $P(\pi_0)$, and $P(\mu,\Sigma)$ facilitates the approximation of the posterior distribution by means of variational Bayes inference \cite{Rezek2002}. To this end, the posterior distribution is approximated to factorise, such that
\begin{align}
   P(y,s,\pi_0,\pi,\mu_k,\Sigma_k) \approx P(y)P(s)P(\pi_0,\pi)P(\mu,\sigma) =\vcentcolon Q.
\end{align}
Q is then determined by minimising the variational free energy \cite{Bishop2013} between the true posterior and this approximation. Up to an additive constant this free energy equals the KL divergence between the two distributions.

\subsection{Application to EEG envelopes}
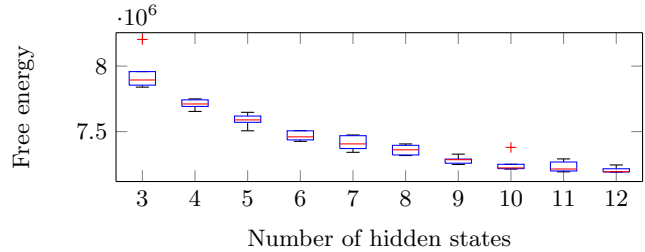
\begin{figure}[htpb]
	\centering
	\setlength\figureheight{2cm} 
	\setlength\figurewidth{.5\linewidth}
	\input{./free_energy_eeg.tikz}
        \caption{\textbf{Free energy of the variational approximation} -- Variational inference is repeated 10 times for every fixed number of Markov states.\label{fig:free_energy}}
 \renewcommand{\thesubfigure}{\alph{subfigure}} \end{figure}

The number of states that is supported by the EEG data is investigated by plotting the free energy as a function of the number of states, as shown in Fig.~\ref{fig:free_energy}.
The free energy decreases steadily for larger numbers of states, which is in agreement with earlier observations by Baker and colleagues~\cite{Baker2014} on resting state MEG data. However, we base our choice of the number of states onto comparability with classical EEG microstate analysis (4 states).

The HMM is applied to the 20 first principal components of the EEG power envelopes. State topographies are derived with a general linear model with the inferred HMM state time course as design matrix and the EEG sensor space power as response. The resulting coefficients are maps of partial correlations, shown in Fig.~\ref{fig:eeg_topos}~\subref{subfig:hmm_topos}. 

\section{Comparison of EEG microstates and Markov states}
\label{sec:comp-eeg-micr}
EEG microstates and EEG HMM states show some spatial similarities (Fig.~\ref{fig:eeg_topos}). The activation is mostly limited to one specific region of sensor space and both microstate and HMM state topographies can be very broadly classified as right lateral, left lateral, anterior/central, and posterior. The lateral microstates expand more into anterior areas, while the corresponding HMM states are laterally confined. Notably, the absolute range of potential differences differs between HMM states and is virtually identical between microstates.

We further compare the temporal properties of both sets of states. Similarly to the HMM state analysis, microstate time courses are obtained as partial correlation between each microstate topography and the EEG power envelope time course. The most probable state $s_t$ at every time point is derived by the Viterbi algorithm \cite{Rezek2002}. It facilitates an estimate of the overall fractional occupancy of each state, which is similar between states and models. The difference in the time spent in two corresponding states is 15\% or less. 

\begin{figure}[b]
\centering
\subfloat[\label{fig:hmm_ms_comp} Correlation structure of microstate and HMM state time courses]{
\includegraphics[height=.4\linewidth]{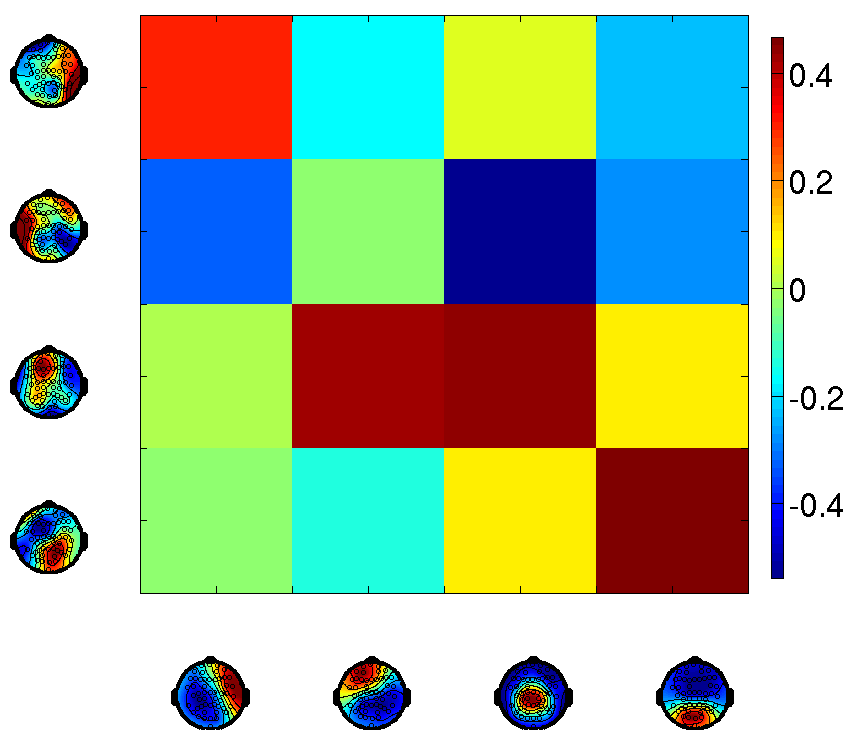} 
} \hspace{.03\linewidth} %
\subfloat[\label{fig:hmm_ms_spatial_comp} Correlation structure of microstate and HMM state spatial maps]{
\includegraphics[height=.4\linewidth]{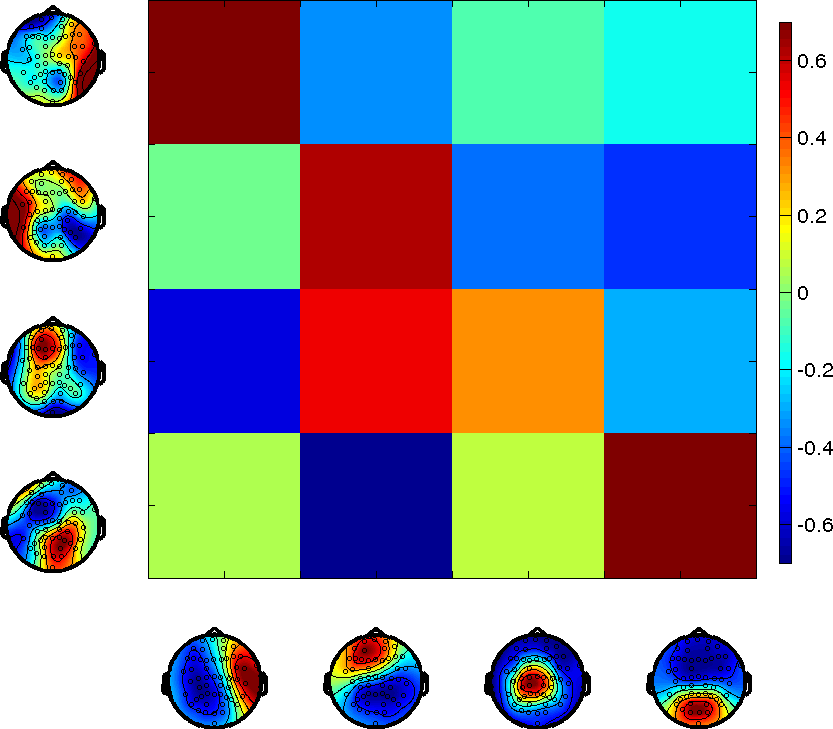} 
}
\caption{Spatial and temporal correlation structure of microstates and HMM state\label{fig:corr_structure}}
\end{figure}

Pairwise correlations of the full time courses are shown in Fig.~\ref{fig:corr_structure}~\subref{fig:hmm_ms_comp}. 
The corresponding spatial correlations are shown in Fig.~\ref{fig:corr_structure}~\subref{fig:hmm_ms_spatial_comp}. Overall, they exhibit a very similar pattern. There are however clear difference, as for instance a strong spatial correlation between microstate (b) and HMM state (b), which is not reflect in the temporal correlation. Conversely microstate (d) and HMM state (c) feature a moderately positive time course correlation, while the spatial patterns are negatively correlated.
Since the time courses for the two modalities are partial correlations of the topographies with the identical envelope time courses it seems unsurprising that temporal correlations are in close agreement with the spatial overlap of the topographies. However, it is clear that the temporal evolution features information complimentary to the 2D topographies.  

\begin{figure}[tp]
\centering
\includegraphics[width=\linewidth]{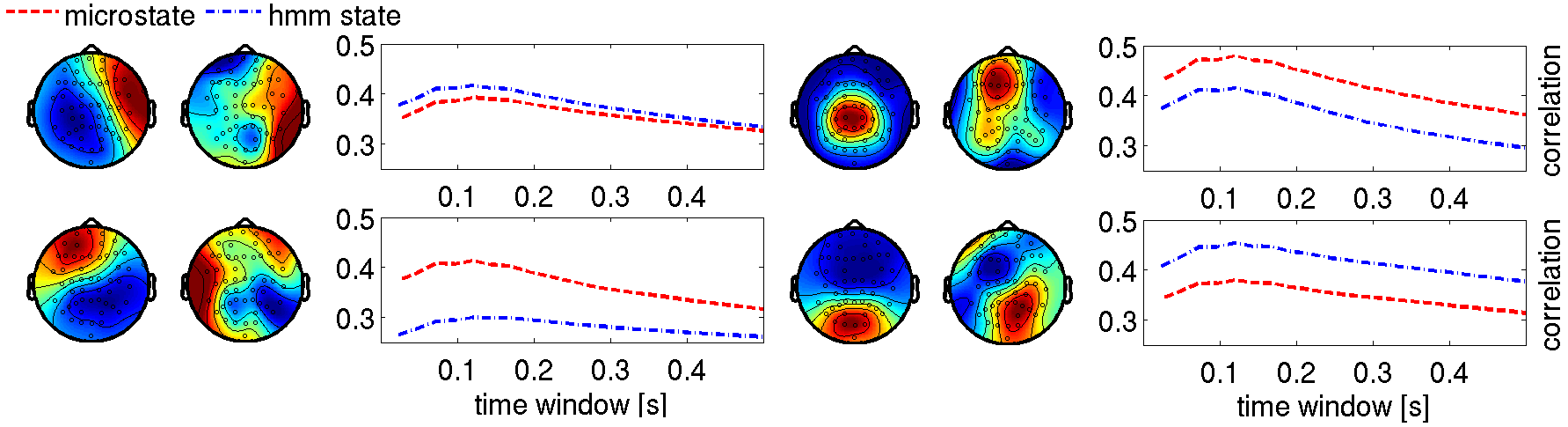}
\caption{\label{fig:low_pass_corr} \textbf{Time scale analysis of within-topography fluctuations} -- The fractional occupancy time window dependency of the correlation between the microstate and HMM state time course and the envelope of the most representative EEG sensor is shown and exhibits consistent maxima around 100--150\,ms}
\end{figure}

To investigate the time scale of the inferred dynamics of state switching as supported by the envelope data we correlate low pass filtered versions of each state time course with the envelope fluctuation of a representative EEG sensor. We selected the sensor that had the highest correlation with the unfiltered state time course and repeated this analysis for every microstate and every HMM state varying the width of the filter.
The results are shown in Fig.~\ref{fig:low_pass_corr} and consistently exhibit maxima in the correlation at window width of 100--150\,ms. 
This agrees with previous observations for EEG microstates \cite{Lehmann1987}. None of the modalities shows a steadily higher correlation than the other.
\begin{figure*}[ptb]
\centering
\subfloat[HMM state time course correlation structure \label{subfig:hmm_corr_struct}]{%
\includegraphics[height=.4\linewidth]{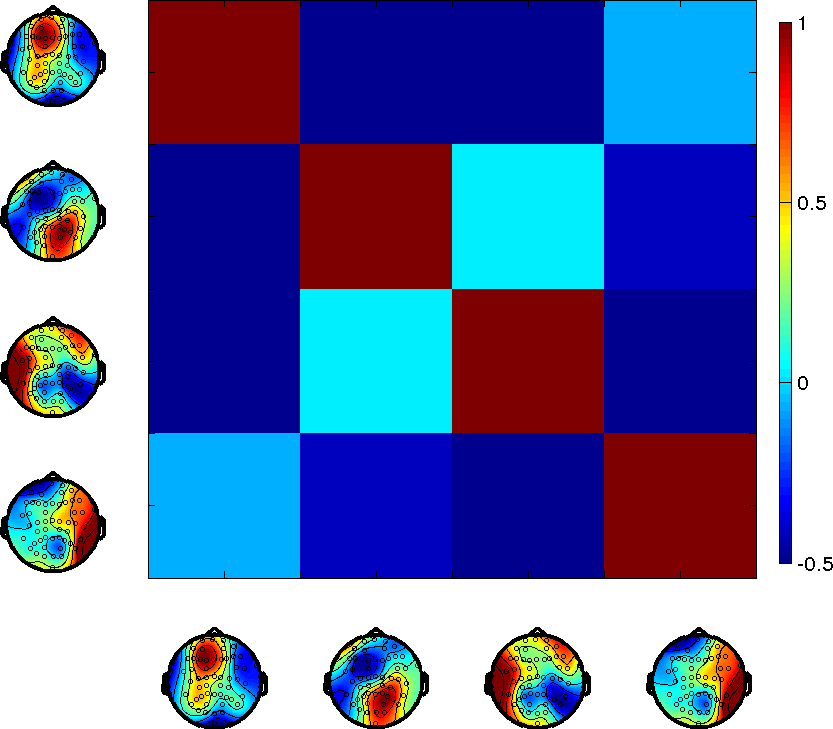} 
} \hspace{.05\linewidth}
\subfloat[Microstate time course correlation structure \label{subfig:ms_corr_struct}]{%
\includegraphics[height=.4\linewidth]{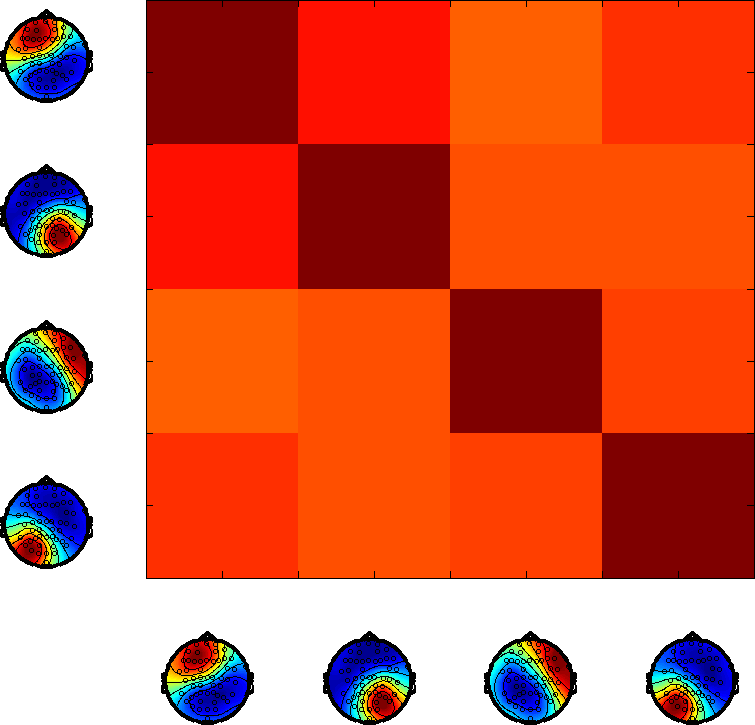}
}
\caption{\label{fig:hmm_ms_corr} \textbf{Correlation structure of the state time courses for the different modalities} -- A strong positive correlation between all EEG microstate time courses is visible, while different HMM state time courses are approximately uncorrelated or negatively correlated.}
\end{figure*}
Interestingly, we observe a very different correlations between the microstate time courses as shown in comparison to the HMM time courses in Fig.~\ref{fig:hmm_ms_corr}.
The microstate time courses exhibit pairwise correlations that are consistently larger than 0.6. Conversely the HMM state time courses show correlations down to -0.5. 

\section{Discussion and conclusion}
\label{sec:disc-concl}
We identified states of quasi-stable topographies in resting state EEG by means of classical microstate analysis and proposed an alternative approach based on a hidden Markov model with a Gaussian observation model that was tractable for approximate inference using Variational Bayes~\cite{Vidaurre2015_unp}. The microstate analysis identified topographies that are similar to known microstates~\cite{Khanna2015}. 
While the HMM state topographies are less confined, they feature similar activation patterns and can partly be matched to corresponding microstates. This matching was shown to be reflected in the temporal evolution of the state time courses. Notably the absolute values of these correlations reach their maxima at about $0.4$, pointing to a dissimilarity that may correspond to a loss/gain of meaningful information in one of the methods.
The EEG microstate time courses show strong pairwise temporal correlations, which is not observed for EEG HMM states. While a temporal (and spatial) overlap between RSNs is entirely possible, networks of different function should also be temporally distinct \cite{Smith2012a}. Thus, a high positive correlation between the dynamics of all given states is unlikely to be a good description of underlying physiology.
We find dominating state lifetimes of 100--150\,ms, which is consistent with earlier findings \cite{Lehmann1987,Baker2014}.

The free energy, a measure for the HMM model fitness, decreases steadily for an increase in the number of states. A possible explanation for the absence of an optimal number of states within the investigated range is the following. For a higher number of states, subject specific activations are introduced in addition to the desirable patterns that are present across subjects. We frequently observe such subject specific states when increasing the number of states (not shown here). While this was partly amendable by demeaning and variance normalising the subject-wise power envelopes, the reason for this behaviour is likely the actually distinct covariance structure between subjects. To make the HMM analysis robust and reliable, this issue should be addressed in future work. More generally, a nonparametric model could automatically infer the optimal number of states.

Further work should include the investigation of the scaling behaviour of HMM states for both EEG and MEG measurements to ascertain whether they exhibit the long-range dependencies (LRDs) that were found for EEG microstates \cite{VandeVille2010} and that are hypothesised to be necessary for the efficient execution of cognitive processes \cite{Chialvo2008,Haimovici2013,Tagliazucchi2012}. Recently Gschwind et. al~\cite{Gschwind2015} criticised the lack of an explicit representation of LRDs in the context of different model~\cite{Gaertner2015}. The HMM does not explicitly model LRDs and therefore the same criticism applies. However, even without explicitly modelling them, we expect the HMM to capture LRDs, if they are present in the data. Nevertheless, a model that takes explicit account of LRDs could constitute a useful expansion of the HMM.

Overall our results suggest, that EEG HMM states could serve as a model-based, physiologically motivated alternative to classical EEG microstates. However, further work remains to be done to substantiate this proposition and to better understand the relationship between between the two approaches.

\printbibliography

\end{document}

%% file: free_energy_eeg.tikz
%
%
\begin{tikzpicture}

\begin{axis}[%
width=\figurewidth,
height=0.99\figureheight,
at={(0\figurewidth,0\figureheight)},
scale only axis,
unbounded coords=jump,
clip=false,
separate axis lines,
every outer x axis line/.append style={black},
every x tick label/.append style={font=\color{black}},
xmin=0.5,
xmax=10.5,
xtick={1,2,3,4,5,6,7,8,9,10},
xticklabels={3,4,5,6,7,8,9,10,11,12},
xlabel={Number of hidden states},
every outer y axis line/.append style={black},
every y tick label/.append style={font=\color{black}},
ymin=7119190.44211397,
ymax=8254753.24225094,
ylabel={Free energy},
axis background/.style={fill=white}
]
\addplot [color=black,dashed,forget plot]
  table[row sep=crcr]{%
1	7957959.37438808\\
1	7957959.37438808\\
};
\addplot [color=black,dashed,forget plot]
  table[row sep=crcr]{%
2	7742380.65021301\\
2	7749764.43146394\\
};
\addplot [color=black,dashed,forget plot]
  table[row sep=crcr]{%
3	7618774.7901139\\
3	7647219.51498794\\
};
\addplot [color=black,dashed,forget plot]
  table[row sep=crcr]{%
4	7506164.09816397\\
4	7507364.40500737\\
};
\addplot [color=black,dashed,forget plot]
  table[row sep=crcr]{%
5	7468949.29511279\\
5	7474429.91558093\\
};
\addplot [color=black,dashed,forget plot]
  table[row sep=crcr]{%
6	7395369.83420777\\
6	7406027.11942614\\
};
\addplot [color=black,dashed,forget plot]
  table[row sep=crcr]{%
7	7290467.50504395\\
7	7328221.57322234\\
};
\addplot [color=black,dashed,forget plot]
  table[row sep=crcr]{%
8	7250746.09089251\\
8	7250746.09089251\\
};
\addplot [color=black,dashed,forget plot]
  table[row sep=crcr]{%
9	7268099.46646508\\
9	7291738.81769106\\
};
\addplot [color=black,dashed,forget plot]
  table[row sep=crcr]{%
10	7216625.79038699\\
10	7245583.7130266\\
};
\addplot [color=black,dashed,forget plot]
  table[row sep=crcr]{%
1	7839376.43851523\\
1	7854450.93910863\\
};
\addplot [color=black,dashed,forget plot]
  table[row sep=crcr]{%
2	7654035.26892918\\
2	7692218.40737828\\
};
\addplot [color=black,dashed,forget plot]
  table[row sep=crcr]{%
3	7506318.14969029\\
3	7570950.13364295\\
};
\addplot [color=black,dashed,forget plot]
  table[row sep=crcr]{%
4	7425613.14958641\\
4	7436871.94621411\\
};
\addplot [color=black,dashed,forget plot]
  table[row sep=crcr]{%
5	7342104.29580838\\
5	7370895.31969902\\
};
\addplot [color=black,dashed,forget plot]
  table[row sep=crcr]{%
6	7317345.9216254\\
6	7321863.71304966\\
};
\addplot [color=black,dashed,forget plot]
  table[row sep=crcr]{%
7	7249245.24993485\\
7	7259050.45076497\\
};
\addplot [color=black,dashed,forget plot]
  table[row sep=crcr]{%
8	7213887.27302357\\
8	7217557.72972281\\
};
\addplot [color=black,dashed,forget plot]
  table[row sep=crcr]{%
9	7193092.87756067\\
9	7199618.16828757\\
};
\addplot [color=black,dashed,forget plot]
  table[row sep=crcr]{%
10	7190827.07780043\\
10	7191828.74191099\\
};
\addplot [color=black,solid,forget plot]
  table[row sep=crcr]{%
0.875	7957959.37438808\\
1.125	7957959.37438808\\
};
\addplot [color=black,solid,forget plot]
  table[row sep=crcr]{%
1.875	7749764.43146394\\
2.125	7749764.43146394\\
};
\addplot [color=black,solid,forget plot]
  table[row sep=crcr]{%
2.875	7647219.51498794\\
3.125	7647219.51498794\\
};
\addplot [color=black,solid,forget plot]
  table[row sep=crcr]{%
3.875	7507364.40500737\\
4.125	7507364.40500737\\
};
\addplot [color=black,solid,forget plot]
  table[row sep=crcr]{%
4.875	7474429.91558093\\
5.125	7474429.91558093\\
};
\addplot [color=black,solid,forget plot]
  table[row sep=crcr]{%
5.875	7406027.11942614\\
6.125	7406027.11942614\\
};
\addplot [color=black,solid,forget plot]
  table[row sep=crcr]{%
6.875	7328221.57322234\\
7.125	7328221.57322234\\
};
\addplot [color=black,solid,forget plot]
  table[row sep=crcr]{%
7.875	7250746.09089251\\
8.125	7250746.09089251\\
};
\addplot [color=black,solid,forget plot]
  table[row sep=crcr]{%
8.875	7291738.81769106\\
9.125	7291738.81769106\\
};
\addplot [color=black,solid,forget plot]
  table[row sep=crcr]{%
9.875	7245583.7130266\\
10.125	7245583.7130266\\
};
\addplot [color=black,solid,forget plot]
  table[row sep=crcr]{%
0.875	7839376.43851523\\
1.125	7839376.43851523\\
};
\addplot [color=black,solid,forget plot]
  table[row sep=crcr]{%
1.875	7654035.26892918\\
2.125	7654035.26892918\\
};
\addplot [color=black,solid,forget plot]
  table[row sep=crcr]{%
2.875	7506318.14969029\\
3.125	7506318.14969029\\
};
\addplot [color=black,solid,forget plot]
  table[row sep=crcr]{%
3.875	7425613.14958641\\
4.125	7425613.14958641\\
};
\addplot [color=black,solid,forget plot]
  table[row sep=crcr]{%
4.875	7342104.29580838\\
5.125	7342104.29580838\\
};
\addplot [color=black,solid,forget plot]
  table[row sep=crcr]{%
5.875	7317345.9216254\\
6.125	7317345.9216254\\
};
\addplot [color=black,solid,forget plot]
  table[row sep=crcr]{%
6.875	7249245.24993485\\
7.125	7249245.24993485\\
};
\addplot [color=black,solid,forget plot]
  table[row sep=crcr]{%
7.875	7213887.27302357\\
8.125	7213887.27302357\\
};
\addplot [color=black,solid,forget plot]
  table[row sep=crcr]{%
8.875	7193092.87756067\\
9.125	7193092.87756067\\
};
\addplot [color=black,solid,forget plot]
  table[row sep=crcr]{%
9.875	7190827.07780043\\
10.125	7190827.07780043\\
};
\addplot [color=blue,solid,forget plot]
  table[row sep=crcr]{%
0.75	7854450.93910863\\
0.75	7957959.37438808\\
1.25	7957959.37438808\\
1.25	7854450.93910863\\
0.75	7854450.93910863\\
};
\addplot [color=blue,solid,forget plot]
  table[row sep=crcr]{%
1.75	7692218.40737828\\
1.75	7742380.65021301\\
2.25	7742380.65021301\\
2.25	7692218.40737828\\
1.75	7692218.40737828\\
};
\addplot [color=blue,solid,forget plot]
  table[row sep=crcr]{%
2.75	7570950.13364295\\
2.75	7618774.7901139\\
3.25	7618774.7901139\\
3.25	7570950.13364295\\
2.75	7570950.13364295\\
};
\addplot [color=blue,solid,forget plot]
  table[row sep=crcr]{%
3.75	7436871.94621411\\
3.75	7506164.09816397\\
4.25	7506164.09816397\\
4.25	7436871.94621411\\
3.75	7436871.94621411\\
};
\addplot [color=blue,solid,forget plot]
  table[row sep=crcr]{%
4.75	7370895.31969902\\
4.75	7468949.29511279\\
5.25	7468949.29511279\\
5.25	7370895.31969902\\
4.75	7370895.31969902\\
};
\addplot [color=blue,solid,forget plot]
  table[row sep=crcr]{%
5.75	7321863.71304966\\
5.75	7395369.83420777\\
6.25	7395369.83420777\\
6.25	7321863.71304966\\
5.75	7321863.71304966\\
};
\addplot [color=blue,solid,forget plot]
  table[row sep=crcr]{%
6.75	7259050.45076497\\
6.75	7290467.50504395\\
7.25	7290467.50504395\\
7.25	7259050.45076497\\
6.75	7259050.45076497\\
};
\addplot [color=blue,solid,forget plot]
  table[row sep=crcr]{%
7.75	7217557.72972281\\
7.75	7250746.09089251\\
8.25	7250746.09089251\\
8.25	7217557.72972281\\
7.75	7217557.72972281\\
};
\addplot [color=blue,solid,forget plot]
  table[row sep=crcr]{%
8.75	7199618.16828757\\
8.75	7268099.46646508\\
9.25	7268099.46646508\\
9.25	7199618.16828757\\
8.75	7199618.16828757\\
};
\addplot [color=blue,solid,forget plot]
  table[row sep=crcr]{%
9.75	7191828.74191099\\
9.75	7216625.79038699\\
10.25	7216625.79038699\\
10.25	7191828.74191099\\
9.75	7191828.74191099\\
};
\addplot [color=red,solid,forget plot]
  table[row sep=crcr]{%
0.75	7893993.21676767\\
1.25	7893993.21676767\\
};
\addplot [color=red,solid,forget plot]
  table[row sep=crcr]{%
1.75	7710926.94334331\\
2.25	7710926.94334331\\
};
\addplot [color=red,solid,forget plot]
  table[row sep=crcr]{%
2.75	7589436.91631514\\
3.25	7589436.91631514\\
};
\addplot [color=red,solid,forget plot]
  table[row sep=crcr]{%
3.75	7460046.35451229\\
4.25	7460046.35451229\\
};
\addplot [color=red,solid,forget plot]
  table[row sep=crcr]{%
4.75	7406272.64400233\\
5.25	7406272.64400233\\
};
\addplot [color=red,solid,forget plot]
  table[row sep=crcr]{%
5.75	7361300.47541655\\
6.25	7361300.47541655\\
};
\addplot [color=red,solid,forget plot]
  table[row sep=crcr]{%
6.75	7285753.48096242\\
7.25	7285753.48096242\\
};
\addplot [color=red,solid,forget plot]
  table[row sep=crcr]{%
7.75	7223130.51794079\\
8.25	7223130.51794079\\
};
\addplot [color=red,solid,forget plot]
  table[row sep=crcr]{%
8.75	7214700.00849754\\
9.25	7214700.00849754\\
};
\addplot [color=red,solid,forget plot]
  table[row sep=crcr]{%
9.75	7194428.0977647\\
10.25	7194428.0977647\\
};
\addplot [color=blue,only marks,mark=+,mark options={solid,draw=red},forget plot]
  table[row sep=crcr]{%
1	8203136.75133562\\
};
\addplot [color=blue,only marks,mark=+,mark options={solid,draw=red},forget plot]
  table[row sep=crcr]{%
nan	nan\\
};
\addplot [color=blue,only marks,mark=+,mark options={solid,draw=red},forget plot]
  table[row sep=crcr]{%
nan	nan\\
};
\addplot [color=blue,only marks,mark=+,mark options={solid,draw=red},forget plot]
  table[row sep=crcr]{%
nan	nan\\
};
\addplot [color=blue,only marks,mark=+,mark options={solid,draw=red},forget plot]
  table[row sep=crcr]{%
nan	nan\\
};
\addplot [color=blue,only marks,mark=+,mark options={solid,draw=red},forget plot]
  table[row sep=crcr]{%
nan	nan\\
};
\addplot [color=blue,only marks,mark=+,mark options={solid,draw=red},forget plot]
  table[row sep=crcr]{%
nan	nan\\
};
\addplot [color=blue,only marks,mark=+,mark options={solid,draw=red},forget plot]
  table[row sep=crcr]{%
8	7379626.91448809\\
};
\addplot [color=blue,only marks,mark=+,mark options={solid,draw=red},forget plot]
  table[row sep=crcr]{%
nan	nan\\
};
\addplot [color=blue,only marks,mark=+,mark options={solid,draw=red},forget plot]
  table[row sep=crcr]{%
nan	nan\\
};
\end{axis}
\end{tikzpicture}%